\documentclass[a4paper, 10pt, conference]{ieeeconf}      % Use this line for a4 paper

\IEEEoverridecommandlockouts                              % This command is only needed if 
\usepackage{cite}
\usepackage{amsmath,amssymb,amsfonts}
\usepackage{algorithmic}
\usepackage{graphicx}
\usepackage{textcomp}
\usepackage{xcolor}
\usepackage{multirow}
\usepackage{glossaries}
\usepackage{tikz}
\usetikzlibrary{arrows,calc}
\usepackage{pgfplots}
\usepackage[binary-units]{siunitx}
\usepackage{url}

\usepackage{breakurl}
\usepackage{varwidth}
\usepackage{caption}
\usepgfplotslibrary{groupplots}
\usetikzlibrary{shapes.misc, shapes.geometric}
\tikzset{cross/.style={cross out, draw, 
		minimum size=2*(#1-\pgflinewidth), 
		inner sep=0pt, outer sep=0pt, very thick}}
\usetikzlibrary{arrows.meta}
\usetikzlibrary{calc,patterns,angles,quotes}
\newenvironment{customlegend}[1][]{%
	\begingroup
	\csname pgfplots@init@cleared@structures\endcsname
	\pgfplotsset{#1}%
}{%
	\csname pgfplots@createlegend\endcsname
	\endgroup
}%
\usetikzlibrary{external}
\usepackage{textcomp}
\usepackage{lipsum}
\newcommand\copyrighttext{%
	\footnotesize \textcopyright 2020 IEEE. Personal use of this material is permitted. Permission from IEEE must be obtained for all other uses, in any current or future media, including reprinting/republishing this material for advertising or promotional purposes, creating new collective works, for resale or redistribution to servers or lists, or reuse of any copyrighted component of this work in other works.
}
\newcommand\copyrightnotice{%
	\tikzset{external/export=false}
	\begin{tikzpicture}[remember picture,overlay]
	\node[anchor=south,yshift=10pt, xshift=10pt] at (current page.south) {\fbox{\parbox{\dimexpr\textwidth-\fboxsep-\fboxrule\relax}{\copyrighttext}}};
	\end{tikzpicture}%
	\tikzset{external/export=true}
}

\definecolor{colBlue}{RGB}{0,101,189}%
\definecolor{colBlueDark}{RGB}{0,82,147}%
\definecolor{colRed}{RGB}{227,114,34}%
\definecolor{colGreen}{RGB}{162,173,0}%
\definecolor{colGray}{RGB}{153,153,153}%
\definecolor{mycolor1}{rgb}{0.85000,0.32500,0.09800}%
\definecolor{mycolor2}{rgb}{0.92900,0.69400,0.12500}%
\definecolor{TUMgray}{rgb}{0.5977,0.5977,0.5977}
\definecolor{TUMgrayLight}{rgb}{0.8516,0.8398,0.7929}

\def\addlegendimage{\csname pgfplots@addlegendimage\endcsname}

\newacronym{EMS}{EMS}{Energy Management Strategy}
\newacronym{DC}{DC}{Direct Current}
\newacronym{AC}{AC}{Alternating Current}
\newacronym{OCP}{OCP}{Optimal Control Problem}
\newacronym{SOCP}{SOCP}{Second Order Conic Problems}
\newacronym{SQP}{SQP}{Sequential Quadratic}
\newacronym{NLP}{NLP}{Nonlinear Problem}
\newacronym{MLTP}{MLTP}{Minimum Lap Time Problem}
\newacronym{MPC}{MPC}{Model Predictive Control}
\newacronym{SOC}{SOC}{State of Charge}
\newacronym{TUM}{TUM}{Technical University of Munich}
\newacronym{ODE}{ODE}{Ordinary Differential Equations}
\newacronym{HIL}{HiL}{Hardware-in-the-Loop}
\newacronym{ICE}{ICE}{Internal Combustion Engine}
\newacronym{MSE}{MSE}{Mean Square Error}

\title{\LARGE \bf
Minimum Race-Time Planning-Strategy\\for an Autonomous Electric Racecar
}

\author{Thomas Herrmann$^{1}$, Francesco Passigato$^{1}$, Johannes Betz$^{1}$ and Markus Lienkamp$^{1}$% <-this % stops a space
\thanks{$^{1}$Thomas Herrmann (corresponding author), Francesco Passigato, Johannes Betz and Markus Lienkamp are with the Chair of Automotive Technology, Faculty of Mechanical Engineering, Technical University of Munich, 85748 Garching (Munich), Germany
        {\tt\small thomas.herrmann@tum.de}}%
}

\begin{document}

\maketitle
\copyrightnotice
\thispagestyle{empty}
\pagestyle{empty}

%%%%%%%%%%%%%%%%%%%%%%%%%%%%%%%%%%%%%%%%%%%%%%%%%%%%%%%%%%%%%%%%%%%%%%%%%%%%%%%%
\begin{abstract}
Increasing attention to autonomous passenger vehicles has also attracted interest in an autonomous racing series. Because of this, platforms such as Roborace and the Indy Autonomous Challenge are currently evolving.\\
Electric racecars face the challenge of a limited amount of stored energy within their batteries. Furthermore, the thermodynamical influence of an all-electric powertrain on the race performance is crucial. Severe damage can occur to the powertrain components when thermally overstressed.
In this work we present a race-time minimal control strategy deduced from an \gls{OCP} that is transcribed into a \gls{NLP}. Its optimization variables stem from the driving dynamics as well as from a thermodynamical description of the electric powertrain. We deduce the necessary first-order \gls{ODE}s and form simplified loss models for the implementation within the numerical optimization. The significant influence of the powertrain behavior on the race strategy is shown.
\end{abstract}

%%%%%%%%%%%%%%%%%%%%%%%%%%%%%%%%%%%%%%%%%%%%%%%%%%%%%%%%%%%%%%%%%%%%%%%%%%%%%%%%
\section{INTRODUCTION}
The first autonomous race series for all-electric racecars is called \textit{Roborace}. Its main goal is to be a platform for the development of software powering self-driving cars. Therefore, Roborace is a race format, bringing cars to their limits of handling \cite{Betz2019c}. The special requirements for the algorithms regarding computational resources, real-time capability, and robustness are thus outstanding \cite{Betz2019}, \cite{Betz2019b}. The race format of autonomous motor-sports delivers perfect conditions for testing under tough conditions in an enclosed environment. We, a team from the \gls{TUM}, are participating in this race series. Most parts of our software stack are available online \cite{ChairofAutomotiveTechnology2019}. This paper describes the extension of our software by a control strategy calculating the minimum race-time, taking into account energetic and thermal constraints arising from the powertrain architecture. The minimum time control strategy is one of three parts of our race strategy (Fig. \ref{fig:RaceStrategy}). As discussed in the results section of this paper, the powertrain thermodynamics have a major impact on the entire race strategy.\\
This paper is based on the ideas for an energy management strategy for autonomous electric cars as stated in our previous work \cite{Herrmann2019}. We extend the state of the art by taking into account multiple race laps as well as the thermodynamics of the all-electric powertrain in the \gls{OCP} that needs to be solved.

\subsection{State of the Art}
\label{subsec:StateOfArt}
\gls{OCP}s dealing with trajectory optimization, which is equivalent to solving a \gls{MLTP} in motor-sports, are well known in the literature. Different mathematical approaches are used to solve an \gls{MLTP}. The variety ranges from graph search methods \cite{Jeon2013} over \gls{SQP} \cite{Carvalho2013} to \gls{SOCP} \cite{Ebbesen2018}. By the transformation of the \gls{OCP} into an \gls{NLP}, the \gls{MLTP} can be solved with detailed and complex double-track vehicle and tire models for the purpose of, e.g., vehicle parameter optimization for \gls{ICE} powered cars \cite{Kirches2010,Limebeer2015,ImaniMasouleh2016}. Latest publications in the field of trajectory optimization also consider optimal power distributions within hybrid powertrains \cite{Salazar2017} or use \gls{MPC}-approaches for the planning of energy-saving trajectories \cite{Wu2019}.\\
However, none of these sources consider the thermodynamics of the powertrain during their optimizations. Unless the temperature limits of a conventional \gls{ICE}, the electric machines of the Roborace cars must not reach temperatures beyond \SI{180}{\degreeCelsius}, the inverter's limit is \SI{100}{\degreeCelsius} \cite{Roborace2018}. Additionally, the efficiency level of an electric machine decreases as it heats up, leading to further reduction in efficiency \cite{Schuetzhold2013, Li2013}. Furthermore, the energy storage, a lithium-ion battery in our case, must reduce its output power from \SI{50}{\degreeCelsius} to \SI{0}{\percent} output power when reaching \SI{55}{\degreeCelsius} for safety reasons \cite{Roborace2018}. In order to therefore prevent the unwanted power loss, the thermal behavior of the powertrain components must be considered for consecutive race laps when dealing with all-electric racecars.\\
This paper is organized as follows: In Subsection \ref{subsec:StructureOfRaceStrategy}, an overview of the structure of our race strategy is given. Section \ref{sec:PowertrainAndModeling} describes the powertrain architecture including power loss descriptions, thermodynamical models as well as the formulation of the optimization problem. Section \ref{sec:Results} explains the results in detail. A summary of the presented work is given in Section \ref{sec:ConclusionOutlook}.

\subsection{Structure of the race strategy}
\label{subsec:StructureOfRaceStrategy}
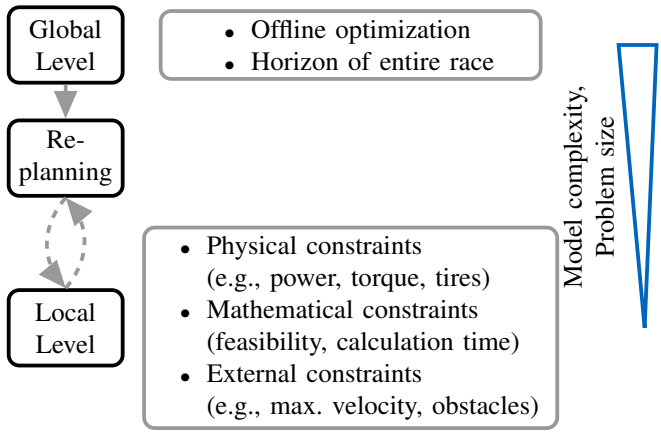
\begin{figure}[h]
	\centering
	\tikzstyle{LevelNode} = [rectangle, rounded corners, minimum width=1.5cm, minimum height=1cm,text centered, draw=black, line width=0.5mm]
\tikzstyle{LevelContent} = [rectangle, rounded corners, minimum width=5cm, minimum height=0.75cm, text centered, draw=colGray, line width=0.5mm]
\tikzstyle{arrow} = [->,-triangle 45, line width=0.5mm,black]
\tikzstyle{lineE} = [-, line width=0.5mm, colGray, dashed]
\tikzstyle{lineM} = [-, line width=1.5mm, black]
\begin{tikzpicture}[node distance=1.25cm]
\coordinate (Zero) at (0,0);
\node (Level0) [LevelNode, align=center] at (Zero) {Global\\Level};
\node (Level2) [LevelNode, align=center, below of=Level0, yshift=-2.5cm] {Local\\Level};
\node (Level1) [LevelNode, align=center, at=($(Level0)!0.4!(Level2)$)] {Re-\\planning};
% Level content
\node (Level0Content) [LevelContent, align=center, right of=Level0, xshift=2.5cm] {
	{\begin{varwidth}{\linewidth}\begin{itemize}
		\item Offline optimization
		\item Horizon of entire race
		\end{itemize}\end{varwidth}}
};
\node (Level2Content) [LevelContent, align=center, right of=Level2, xshift=2.5cm] {
	{\begin{varwidth}{\linewidth}\begin{itemize}
	\item Physical constraints\\(e.g., power, torque, tires)
	\item Mathematical constraints\\(feasibility, calculation time)
	\item External constraints\\(e.g., max. velocity, obstacles)
	\end{itemize}\end{varwidth}}
};
\draw[line width=0.5mm, colBlue] ($(Level0Content.east) + (1,0)$) node[anchor=north]{}
-- ($(Level0Content.east) + (1.5,0)$) node[anchor=north]{}
-- ($(Level2Content.east) + (1.125,0)$) node[anchor=south]{}
-- cycle;

\node[rotate=90] at ($(Level0Content.east)!0.5!(Level2Content.east) + (0.5,0)$) [align=center]{Model complexity,\\Problem size};

% Arrows
\draw [arrow, colGray, dashed]  (Level0.south) to[out=270, in=90] (Level1.north);
\draw [arrow, colGray, dashed]  (Level1.south) to[out=230, in=130, distance=0.5cm] (Level2.north);
\draw [arrow, colGray, dashed]  (Level2.north) to[out=50, in=310, distance=0.5cm] (Level1.south);
\end{tikzpicture}
	\caption{Three levels of the proposed race strategy.}
	\label{fig:RaceStrategy}
\end{figure}

The race strategy is split into three levels. All of these levels have a different optimization horizon as well as a different problem size stemming from the combination of their optimization horizon as well as their model complexities (Fig. \ref{fig:RaceStrategy}).
\\
Before the race starts, the global time-minimal trajectories per lap for the entire race are calculated offline. Here, we can use a non-linear double track model describing the driving dynamics as well as a detailed thermal powertrain model to consider all necessary physical effects in detail. The pre-computed global trajectories are recalculated during the race, reacting to external disturbances, e.g., overtakes, using a significantly reduced optimization problem formulation. For example, the path can often be completely removed during the replanning phase as it is almost equal for all the race laps. The global trajectories are then fed into the local trajectory planner that transforms all given physical constraints from the global trajectory (e.g., max. power, max. torque) as well as mathematical requirements (e.g., guaranteed feasibility, calculation time) into locally optimal paths and velocities \cite{Stahl2019}. Furthermore, the local planner considers external influences, e.g., overtakes opponent cars or reacts to speed limits. In this paper we focus on the global level of the proposed race strategy and describe the offline optimization.

\section{Powertrain architecture \& modeling}
\label{sec:PowertrainAndModeling}

\begin{figure}[h]
	\centering
	\tikzstyle{simpleNode} = [rectangle, rounded corners, minimum width=1cm, minimum height=1cm,text centered, draw=black, line width=0.5mm]
\tikzstyle{arrow} = [->,-triangle 45,line width=0.5mm,black]
\tikzstyle{lineE} = [-,line width=0.5mm,colGray,dashed]
\tikzstyle{lineM} = [-,line width=1.5mm,black]
\tikzstyle{lineR} = [-,line width=1mm,colBlue]
\begin{tikzpicture}[node distance=1cm]
\coordinate (Zero) at (0,0);
\node (Battery) [simpleNode, align=center] {$\mathrm{B}$};
\node (InverterF) [simpleNode, align=center, left of=Battery, xshift=-0.5cm] {${\mathrm{I_{l/r}}}$};
\node (MotorF) [simpleNode, align=center, left of=InverterF,xshift=-0.5cm] {$\mathrm{M_{l/r}}$};
\node (GearF) [simpleNode, align=center, left of=MotorF, xshift=-0.15cm] {$\mathrm{G_{{l/r}}}$};
\node (WheelFR) [simpleNode, align=center, above of=GearF, minimum height=0.6cm, yshift=0.3cm] {$\mathrm{W_{rl}}$};
\node (WheelFL) [simpleNode, align=center, below of=GearF, minimum height=0.6cm, yshift=-0.3cm] {$\mathrm{W_{rr}}$};
\node (Aux) [simpleNode, align=center, above of=Battery, yshift=0.3cm] {$\mathrm{A_x}$};
\node (InverterR) [simpleNode, minimum width=0cm, right of=Battery, draw=white, xshift=0.25cm] {};
\node (RadiatorMI) [simpleNode, align=center, minimum height=0.75cm, below of=InverterF, draw=colBlue, xshift=-0.75cm, yshift=-0.3cm] {$\mathrm{R_{MI}}$};
\node (RadiatorB) [simpleNode, align=center, minimum height=0.75cm, below of=Battery, draw=colBlue, yshift=-0.3cm] {${\mathrm{R_B}}$};
%%% Edges %%%
\draw [lineE] ([yshift=0.2cm]Battery.west) to ([yshift=0.2cm]InverterF.east);
\draw [lineE] ([yshift=-0.2cm]Battery.west) to ([yshift=-0.2cm]InverterF.east);
\draw [lineE] ([yshift=0.2cm]Battery.east) to ([yshift=0.2cm]InverterR.west);
\draw [lineE] ([yshift=-0.2cm]Battery.east) to ([yshift=-0.2cm]InverterR.west);
% To Aux
\draw [lineE] ([xshift=0.2cm]Battery.north) to ([xshift=0.2cm]Aux.south);
\draw [lineE] ([xshift=-0.2cm]Battery.north) to ([xshift=-0.2cm]Aux.south);
\draw [lineE] (InverterF)[anchor=left] to (MotorF)[anchor=right];
\draw [lineE] ([yshift=0.25cm]InverterF.west) to ([yshift=0.25cm]MotorF.east);
\draw [lineE] ([yshift=-0.25cm]InverterF.west) to ([yshift=-0.25cm]MotorF.east);
\draw [lineM] (MotorF)[anchor=left] to (GearF)[anchor=right];
\draw [lineM] (GearF)[anchor=left] to (WheelFR)[anchor=right];
\draw [lineM] (GearF)[anchor=left] to (WheelFL)[anchor=right];
%\draw [arrow] (Battery) to[bend left] (InverterF);
\draw [thick] ($(InverterR.west)+(0,1)$) to [out=300,in=120] ($(InverterR.west)+(0,-1)$);
% RadiatorMI
\draw [lineR] (InverterF)[anchor=bottom] -- ++(0,-0.5) |- (RadiatorMI)[anchor=right];
% dot
\node (IRdot) at ($(InverterF) + (0,-0.8)$) [label={[yshift=-1.6cm]$T_\mathrm{F1,I}$}] {};
\draw[colBlue,fill=colBlue] (IRdot) circle (2.5pt);
\draw [lineR] (RadiatorMI)[anchor=left] -| (MotorF)[anchor=bottom];
% dot
\node (IMdot) at ($(MotorF) + (0,-0.8)$) [label={[yshift=-1.6cm]$T_\mathrm{F1,R_{MI}}$}] {};
\draw[colBlue,fill=colBlue] (IMdot) circle (2.5pt);
\draw [lineR] (MotorF)[anchor=top] -- ++(0,1) -| (InverterF)[anchor=top];
% dot
\node (MIdot) at ($(MotorF)!0.5!(InverterF) + (0,1)$) [label={[yshift=0cm]$T_\mathrm{F1,M}$}] {};
\draw[colBlue,fill=colBlue] (MIdot) circle (2.5pt);
%
% Radiator B
\draw [lineR] ([xshift=0.2cm]RadiatorB.north) to ([xshift=0.2cm]Battery.south);
\draw [lineR] ([xshift=-0.2cm]RadiatorB.north) to ([xshift=-0.2cm]Battery.south);
% dot
\node (Bldot) at ($(Battery)!0.5!(RadiatorB) - (0.2,0.05)$) [label={[yshift=-1.7cm, xshift=-0.1cm]$T_\mathrm{F2,R_\mathrm{B}}$}] {};
\draw[colBlue,fill=colBlue] (Bldot) circle (2.5pt);
% dot
\node (Brdot) at ($(Battery)!0.5!(RadiatorB) + (0.2,-0.05)$) [label={[yshift=-1.7cm, xshift=0.6cm]$T_\mathrm{F2,B}$}] {};
\draw[colBlue,fill=colBlue] (Brdot) circle (2.5pt);
\end{tikzpicture}
	\caption{Electric powertrain architecture of a rear-wheel drive vehicle including two separate cooling circuits.}
	\label{fig:PowertrainRobocar}
\end{figure}
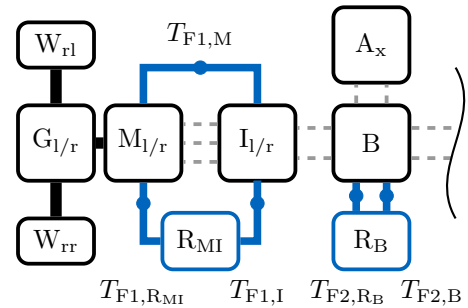

The all-eletric powertrain of the racecar (Fig. \ref{fig:PowertrainRobocar}) consists of
\begin{itemize}
	\item A battery ($\mathrm{B}$).
	\item Two power electronics/inverters at rear left and right ($\mathrm{I_{l/r}}$).
	\item Two synchronous permanent electric machines ($\mathrm{M_{l/r}}$).
\end{itemize}
Two separate cooling circuits are necessary in order to control the component temperatures $T_c$. Circuit 1 is responsible for machines $\mathrm{M_{l/r}}$ and inverters $\mathrm{I_{l/r}}$ leveraging radiator $\mathrm{R_{MI}}$. The same is true for circuit 2, battery $\mathrm{B}$ and radiator $\mathrm{R_{B}}$. For the sake of completeness, the gears ($\mathrm{G_{l/r}}$), sensors required for autonomous driving ($\mathrm{A_x}$), as well as the wheels $\mathrm{W_{rr/rl}}$, are also displayed within the rear part of the whole powertrain.\\
In this work, the index $c$ indicates the powertrain component, i.e., $c~\epsilon~\{\mathrm{M, I, B, R_{MI}, R_B}\}$. The second index $d$ of the temperature symbols of both cooling liquids $T_{\mathrm{F1/2},d}$ enumerates the components the fluids are entering.

\subsection{Power loss models}
\begin{figure}[!h]
	\centering
	\input{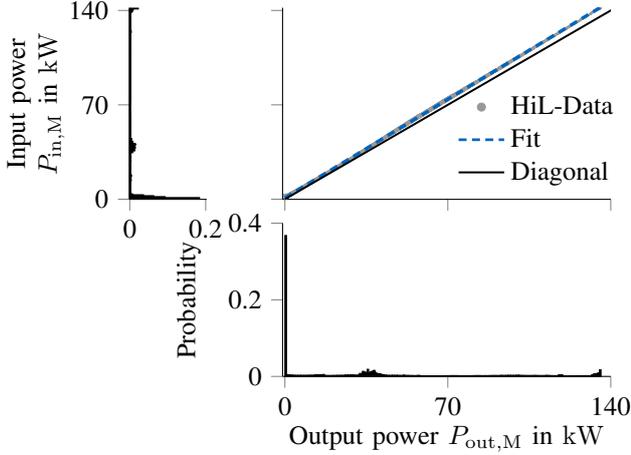}
%	\vspace*{-20mm}  % TODO: adapt space at end of writing process
	\caption{Parabola fit to measurement data to deduce a polynomial expression of the electric machine's efficiency.}
	\label{fig:datafit}
\end{figure}

Meta-models of the powertrain are used to describe the internal losses of its components within the \gls{OCP}. Mathematically, the meta-models for the electric machines as well as the inverters can be formulated as second order polynomials with the output power $P_\mathrm{out}$ as free variable (\ref{eq:ParabolaFit}):
\begin{equation}
	P_{\mathrm{in,fit}}(P_{\mathrm{out}}) = a_{\mathrm{fit}} P_{\mathrm{out}}^2 + b_{\mathrm{fit}} P_{\mathrm{out}} + c_{\mathrm{fit}}
	\label{eq:ParabolaFit}.
\end{equation}
The \gls{MSE} $e_\mathrm{MS,fit}$,
\begin{equation}
	e_\mathrm{MS,fit} = \frac{1}{N} \sum^{N}_{i}{\left( P_{\mathrm{in},\mathrm{fit},i} - P_{\mathrm{in},\mathrm{mes},i} \right)^2}
\end{equation}
is minimized by fitting the constant parameters $a_{\mathrm{fit}}$, $b_{\mathrm{fit}}$, and $c_{\mathrm{fit}}$. The input power $P_{\mathrm{in}}$ into the single components is a function depending on the requested output power $P_{\mathrm{out}}$ \cite{Ebbesen2018}. $P_{\mathrm{in,mes}}$ stems from measurement data from our \gls{HIL}-Simulator \cite{Herrmann2019b} where detailed non-linear powertrain models, that are based on real-world measurement data from the Roborace cars, are implemented. The index $i$ denotes a counter variable in the range $\left[1~..~N\right]$.\\
Fig. \ref{fig:datafit} displays a polynomial fit to simulated data of an electric machine. The probability distributions of data on both axis indicate that mainly max- or minimal power is requested by the racecar. Therefore, the parabolic fit, showing high accuracy at these points, results in low \gls{MSE}s. These are $e_\mathrm{MS,fit,M} = \SI{3.19}{\percent}$ for the electric machine and $e_\mathrm{MS,fit,I} = \SI{4.16}{\percent}$ for the inverter. The diagonal indicates \SI{100}{\percent} efficiency.
\\
We use an open circuit model to describe the internal battery power \cite[p.~51]{Sciarretta2020} $P_\mathrm{in,B}$,
\begin{equation}
	P_{\mathrm{in,B}}(P_{\mathrm{out,B}}) = \frac{U_{\mathrm{OCV}}^2}{2 R_\mathrm{i}} - U_{\mathrm{OCV}} \frac{\sqrt{U_{\mathrm{OCV}}^2 - 4P_{\mathrm{out,B}} R_\mathrm{i}}}{2R_\mathrm{i}}.
	\label{eq:BatteryFit}
\end{equation}
The open circuit voltage is $U_{\mathrm{OCV}}$ and $R_\mathrm{i}$ is the internal battery resistance.\\
The component power loss $P_{\mathrm{los,}c}$ can now be described using (\ref{eq:ParabolaFit}) and (\ref{eq:BatteryFit}) that can easily be implemented within the numerical optimization,
\begin{equation}
	P_{\mathrm{los,}c} = P_{\mathrm{in,}c} - P_{\mathrm{out,}c}.
	\label{eq:PowerLoss}
\end{equation}

\subsection{Thermal models}
\label{subsec:ThermalModels}
As introduced in Section \ref{sec:PowertrainAndModeling}, the thermal model of the powertrain is split into two circuits, one cooling the electric machines $\mathrm{M_{l/r}}$ and both inverters $\mathrm{I_{l/r}}$ in series using the radiator $\mathrm{R_{MI}}$, the other one being responsible for the battery temperature $T_\mathrm{B}$ leveraging radiator $\mathrm{R_{B}}$ (Fig. \ref{fig:PowertrainRobocar}). Power losses (\ref{eq:PowerLoss}) from the electric components $c$ are directly fed into the thermal circuits.\\
In the following, the \gls{ODE}s describing the powertrain thermodynamics, i.e., the heat transfer from the powertrain components to the cooling liquids, are deduced.
%They are similar for every component regarding the heat transfer mechanisms consisting of convection, conduction as well as radiation:
%\begin{align}
%W_{\mathrm{thm}}
%&= \int{ \frac{\left(T_{\mathrm{c}}(t) - T_{\infty}\right)}{R_{\mathrm{cvn}}} \mathrm{d}t}~+\\
%&= \int{ \frac{\left(T_{\mathrm{2}}(t) - T_{\mathrm{1}}(t)\right)}{R_{\mathrm{cdn}}} \mathrm{d}t}~+\\
%&= \int{ \frac{\left(T_{\mathrm{sur}}(t) - T_{\mathrm{sun}}(t)\right)}{R_{\mathrm{rad}}} \mathrm{d}t}.
%\end{align}
In general, the product of the thermal heat capacities $C$ and the temperature gradients $\frac{\mathrm{d}T}{\mathrm{d}t}$ equal loss $P_\mathrm{los}$ and cooling power $P_\mathrm{col}$ for electric machines, inverters and battery:
%
%%% Thermal ODEs
\begin{align}
	\nonumber
	C_\mathrm{M} \frac{\mathrm{d}T_{\mathrm{M}}}{\mathrm{d}t} &= P_{\mathrm{los,M}} - \frac{T_\mathrm{M} - T_\mathrm{M,\infty}}{R_\mathrm{M}} =\\
	&= P_{\mathrm{los,M}} - \underbrace{\frac{2T_\mathrm{M} - \left( T_{\mathrm{F1,M}} + T_{\mathrm{F1,R_{MI}}} \right)}{2R_\mathrm{M}}}_{P_\mathrm{col,M}}\\
	C_\mathrm{I} \frac{\mathrm{d}T_{\mathrm{I}}}{\mathrm{d}t} &= P_{\mathrm{los,I}} - \underbrace{\frac{2T_\mathrm{I} - \left( T_{\mathrm{F1}} + T_{\mathrm{F1,M}} \right)}{2R_\mathrm{I}}}_{P_\mathrm{col,I}}\\
	C_\mathrm{B} \frac{\mathrm{d}T_{\mathrm{B}}}{\mathrm{d}t} &= P_{\mathrm{los,B}} - \underbrace{\frac{T_\mathrm{B} - T_{\mathrm{F2}}}{R_\mathrm{B}}}_{P_\mathrm{col,B}}.
\end{align}
Here and in the following, $T_\mathrm{F1}=T_\mathrm{F1,I}$. The symbol $T_{\mathrm{M},\infty}$ denotes the temperature of the surroundings of the specific component $\mathrm{M}$. This temperature can be assumed to be equal to the mean value of the inflowing and effluent cooling liquid temperature \cite{Han2019}.\\
To describe the absorbed energy by the coolant fluid from both inverters the following formulation is used:
\begin{equation}
	2 \frac{T_{\mathrm{I}}-T_\mathrm{I,\infty}}{R_{\mathrm{I}}} = \dot{m}_{\mathrm{F1}} c_{\mathrm{F}} \left( T_{\mathrm{F1,M}}-T_{\mathrm{F1}} \right),
	\label{eq:qoutI}
\end{equation}
where $\dot{m}_\mathrm{F1}$ describes the coolant mass flow through both inverters and $c_\mathrm{F}$ the specific heat capacity of the coolant fluid. Using $T_\mathrm{{I,\infty}} = \frac{1}{2}\left(T_{\mathrm{F1}} + T_{\mathrm{F1,M}}\right)$, the following equations can be deduced to explicitly describe the temperatures of the cooling liquid entering electric machines as well as the radiator $\mathrm{R_{MI}}$:
\begin{align}
	T_{\mathrm{F1,M}} &= \frac{T_{\mathrm{F1}}\left( \dot{m}_{\mathrm{F1}} c_{\mathrm{F}} R_{\mathrm{I}} - 1 \right) + 2T_{\mathrm{I}}}{1 + \dot{m}_{\mathrm{F1}} c_{\mathrm{F}} R_{\mathrm{I}}}\\
	T_{\mathrm{F1,R_{MI}}} &= \frac{T_{\mathrm{F1}}\left( 2\dot{m}_{\mathrm{F1}} c_{\mathrm{F}} R_{\mathrm{R_{MI}}} + 1 \right) - 2T_{\mathrm{env}}}{2\dot{m}_{\mathrm{F1}} c_{\mathrm{F}} R_{\mathrm{R_{MI}}} - 1}.
	\label{eq:TF1}
\end{align}
%%% Radiator ODEs
We can formulate the gradients of $T_\mathrm{F1}$ and $T_\mathrm{F2}$ using the cooling power of the powertrain components $P_\mathrm{col}$ as well as the temperature differences to the environment $T_\mathrm{env}$,
\begin{align}
	C_\mathrm{F1}\frac{\mathrm{d}T_\mathrm{F1}}{\mathrm{d}t} =&~2 P_{\mathrm{col,M}} + \nonumber\\
	& + 2 P_{\mathrm{col,I}} - \nonumber\\ &- \frac{1}{R_\mathrm{R_{MI}}} \left( \frac{T_\mathrm{F1,R_{MI}} + T_\mathrm{F1}}{2} - T_\mathrm{env} \right) \\
	C_\mathrm{F2}\frac{\mathrm{d}T_\mathrm{F2}}{\mathrm{d}t} =&~P_{\mathrm{col,B}} - \frac{T_\mathrm{F2} - T_\mathrm{env}}{R_\mathrm{R_B}}.
\end{align}
The thermal resistance of the motor model $R_\mathrm{M}$ can be written as a combination of two thermal resistances $R_{1/2}$ in parallel as the heat transfer acts from the air gap to both directions, the environment as well as its shaft (Fig. \ref{fig:MotorThermal}). For this model we make use of \cite{ELRefaie2004} that describes the stator winding temperature $T_\mathrm{W}$ as highest and most critical. $\bar{T}_\mathrm{F1,M}= \frac{1}{2}\left(T_{\mathrm{F1,M}} + T_{\mathrm{F1,R_{MI}}}\right)$ denotes the mean temperature of the cooling liquid through the electric machine. Therefore,
\begin{equation}
	R_\mathrm{M} = \frac{R_1 R_2}{R_1 + R_2},
\end{equation}
with
\begin{align}
	R_1 &= \frac{\ln{\frac{r_4}{r_3}}}{2\pi L k_\mathrm{iro}} + \frac{1}{2 \pi r_4 L h_\mathrm{f}}\\
	R_2 &= \frac{\ln\frac{r_2}{r_1}}{2 \pi L k_\mathrm{iro}} + \frac{1}{4 \pi L k_\mathrm{iro}} + \frac{1}{2 \pi r_3 L h_\mathrm{g}}.
\end{align}
Fig. \ref{fig:MotorThermal} indicates the geometry of the electric machine. The first term in resistance $R_1$ takes into account conduction of the stator where $L$ denotes its length and $k_\mathrm{iro}$ the thermal conductivity of iron. The second term describes the convective heat flux between the stator and the cooling liquid with $h_\mathrm{f}$ being the liquid's convective heat flux coefficient that can be assumed constant \cite[p.~11]{Incropera2007}. Resistance $R_2$ consists of the thermal conductivity of the rotor and shaft \cite{Li2013} as well as the convection into the air gap with the respective heat flux coefficient $h_\mathrm{g}$.
\begin{figure}[h]
	\centering
	\tikzstyle{Circ} = [circle, text centered, draw=black, line width=0.5mm]
\tikzstyle{LevelContent} = [rectangle, rounded corners, minimum width=4cm, minimum height=0.75cm,text centered, draw=colGray, line width=0.5mm]
\tikzstyle{arrow} = [->, -triangle 45, line width=0.3mm, colGray]
\begin{tikzpicture}[node distance=1cm]
\coordinate (Zero) at (0,0);
\node (Shaft) [Circ, align=center, minimum width=2cm, label={[yshift=-0.75cm]Shaft}] at (Zero) {};
\node (Rotor) [Circ, align=center, minimum width=4cm, label={[yshift=-0.75cm]Rotor}] at (Zero) {};
\node (AirGap) [Circ, align=center, minimum width=4.5cm] at (Zero) {};
\node (Stator) [Circ, align=center, minimum width=6cm, label={[yshift=-0.75cm]Stator}] at (Zero) {};
% Level content
%\node (Level0Content) [LevelContent, align=center, right of=Level0, %xshift=2.5cm] {Calculation before race\\Horizon: Entire race};
%%% Edges %%%
\draw [arrow] ($(Shaft.west) + (1,0)$) to (180:1cm) node[label={[xshift=0.5cm, yshift=-0.6cm, black]:$r_1$}]{};
\draw [arrow] ($(Shaft.west) + (1,0)$) to (230:2cm) node[label={[xshift=0.6cm, yshift=-0.2cm, black]:$r_2$}]{};
\draw [arrow] ($(Shaft.west) + (1,0)$) to (280:2.25cm) node[label={[xshift=0.2cm, yshift=0.3cm, black]:$r_3$}]{};
\draw [arrow] ($(Shaft.west) + (1,0)$) to (330:3cm) node[label={[xshift=-0.1cm, yshift=0.1cm, black]:$r_4$}]{};
% Dots
\node (Tshaft) at ($(Shaft)$) [label={[right]$\bar{T}_{\mathrm{F1,M}}$}] {};
\draw[colBlue,fill=colBlue] (Tshaft) circle (2.5pt);
\node (TshaftNE) at ($(Shaft.north east)$) [] {};
\draw[colBlue,fill=colBlue] (TshaftNE) circle (2.5pt);
\node (TrotorNE) at ($(Rotor.north east)$) [] {};
\draw[colBlue,fill=colBlue] (TrotorNE) circle (2.5pt);
\node (TairNE) at ($(AirGap.north east)$) [label={[right]$T_{\mathrm{W}}$}] {};
\draw[colBlue,fill=colBlue] (TairNE) circle (2.5pt);
\node (TstatorNE) at ($(Stator.north east)$) [label={[right]$\bar{T}_{\mathrm{F1,M}}$}] {};
\draw[colBlue,fill=colBlue] (TstatorNE) circle (2.5pt);
% bended arrows
\draw [arrow, colBlue] (TairNE) to[out=280, in=340, distance=1cm] (Tshaft);
\draw [arrow, colBlue] (TairNE) to[out=330, in=300, distance=0.75cm] (TstatorNE);
\begin{customlegend}[legend entries={$P_\mathrm{col,M}$},legend style={at={(2.5,-2.5)}, anchor=north west, align=left, draw=none, fill=none}]
\addlegendimage{arrow,draw=colBlue,fill=none}
\end{customlegend}
\end{tikzpicture}
	\caption{Thermal resistance model of the electric machine.}
	\label{fig:MotorThermal}
\end{figure}
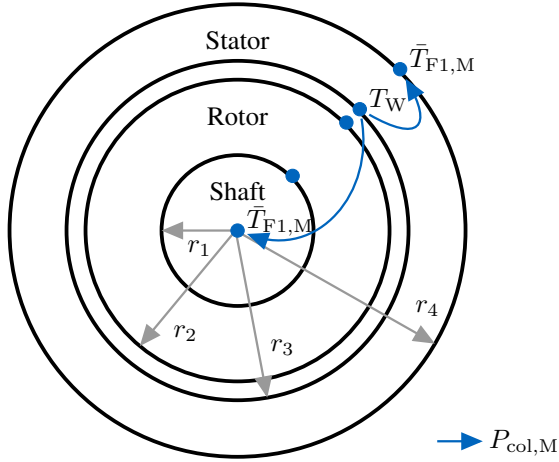\\
The thermal resistance of the inverter is assumed to be
\begin{equation}
	R_\mathrm{I} = \frac{1}{A_\mathrm{I} h_\mathrm{I}},
\end{equation}
as well as for the battery \cite{Han2019}
\begin{equation}
	R_\mathrm{B} = \frac{1}{A_\mathrm{B} h_\mathrm{B}},
\end{equation}
and the radiators
\begin{equation}
	R_\mathrm{R_{MI/B}} = \frac{1}{A_\mathrm{R_{MI/B}} h_\mathrm{R_{MI/B}}}.
\end{equation}
Here, $A$ denotes the surface used for the heat exchange, $h$ again represents heat flux coefficients.
\subsection{Optimization problem}
The \gls{OCP} transformed into an \gls{NLP} with equality and inequality constraints $h_i$ and $g_j$ has the form \cite[p.~478]{Bertsekas2016}, \cite[p.~127]{Boyd2004}
\begin{align}	
	\min~ l(\boldsymbol{x}) &= \int_{0}^{S_{\mathrm{\Sigma}}}{\frac{\mathrm{d}t}{\mathrm{d}s} \mathrm{d}s} = \int_{0}^{S_{\mathrm{\Sigma}}}{\frac{1-n\kappa}{v\cos{\left(\xi + \beta\right)}}\mathrm{d}s}\\
	s.t.~\frac{\mathrm{d}\boldsymbol{x}}{\mathrm{d}s} &= f(\boldsymbol{x}(s), \boldsymbol{u}(s))\\
	h_i &= 0\\
	g_j &\leq 0
	\label{eq:NLPP}
\end{align}
with $i = 1, ..., m$ and $j = 1, ..., r$. The independent variable is $s$, the distance along the reference line, $\kappa$ denotes the curvature profile of the race track. The objective is $l(\boldsymbol{x})$, defined as the integral over the lethargy $\frac{\mathrm{d}t}{\mathrm{d}s}$ \cite{Ebbesen2018} being minimized over the entire racing distance $S_\Sigma$. The lethargy can be interpreted as the time taken to travel a distance of \SI{1}{\meter}.\\
The state vector $\boldsymbol{x}(s)$ within the \gls{OCP} is defined as
\begin{equation}
	\boldsymbol{x}(s) = 
	\left( \underbrace{v~\beta~\dot{\psi}~n~\xi}_{\substack{\text{Driving}\\\text{Dynamics}}}~\underbrace{T_\mathrm{M}~T_\mathrm{I}~T_\mathrm{B}~T_\mathrm{F1}~T_\mathrm{F2}}_{\substack{\text{Thermo-}\\ \text{dynamics}}} \right)^T,
\end{equation}
consisting of the optimization variables needed to express the thermodynamics as introduced in Section \ref{subsec:ThermalModels} as well as the variables defining the driving dynamics. These variables are the velocity $v$ on the raceline, the side slip angle $\beta$, the yaw angle $\psi$, the lateral distance to the reference line $n$, and $\xi$ as the relative angle of the vehicle's longitudinal axis to the tangent on the reference line. For a detailed description of the driving dynamics as well as their first-order \gls{ODE}s and constraints stemming from a nonlinear double track model, we refer to our previous works \cite{Christ2019, Herrmann2019}, as we focus on the thermodynamical side within this paper.\\
The box constraints that are translated into inequality constraints for the thermodynamical state variables are
\begin{equation}
	T_{c,\mathrm{min}} \leq T_c \leq T_{c,\mathrm{max}},
\end{equation}
with every component's allowed operating temperature range defining the lower $T_{c,\mathrm{min}}$ and upper boundaries $T_{c,\mathrm{max}}$.\\
The input vector has the form
\begin{equation}
	\boldsymbol{u}(s) = 
	\begin{pmatrix}
	F_\mathrm{d}~F_\mathrm{b}~\delta~\gamma
	\end{pmatrix}^T,
\end{equation}
containing driving and braking force $F_\mathrm{d}$/$F_\mathrm{b}$, the steering angle $\delta$, and the wheel load transfer $\gamma$ as artificial control variable.\\
%\subsection{Spatial discretization model}
%\label{subsec:SpatialDiscModel}
%In order to solve the optimization problem (high model complexity/complex constraints: NDTM, Thermal Model, high number of optimization variables: Variable Geometry, Variable Velocity). Therefore, strategy to speed up the optimization necessary.\\
%In order to achieve fastest lap times, the full tire potential must be used in the circuit's curves to maximize the lateral acceleration. Therefore, high gradients regarding the optimization variables as well as the constraint formulations, arise. Furthermore, several critical constraints, which ensure the computation of physically feasible trajectories become active, as the full tire potential gets leveraged. In order to deal with this issue, a finer mesh in the curves compared to the straight parts of the circuit is implemented to allow for a better description of the rapidly changing variables. Good numerical precision in combination with small computation times results.\\
%\textcolor{blue}{If space left: put picture of variable discretization length here. Else: rm this section as no evidence for these statements is shown.}
%
%
%%%%%%%%%%%%%%%%%%%%%%%%%%%%%%%%%%%%%%%%%%%%%%%%%%%%%%%%%%%%%%%%%%%%%%%%%%%%%%%%%%%%%%%%%%%%%%%%%%%%%%%%%%%%%%%%
\section{Results}
\label{sec:Results}
\begin{figure}[!h]
	\centering
	\input{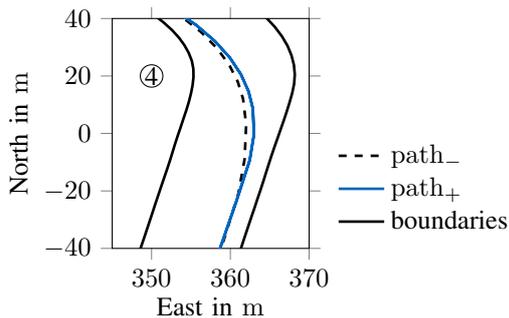}
	\caption{Race paths for case ``cold'' and ``hot''.}
	\label{fig:plot_path}
\end{figure}

\begin{figure*}[h]
	\centering
	\input{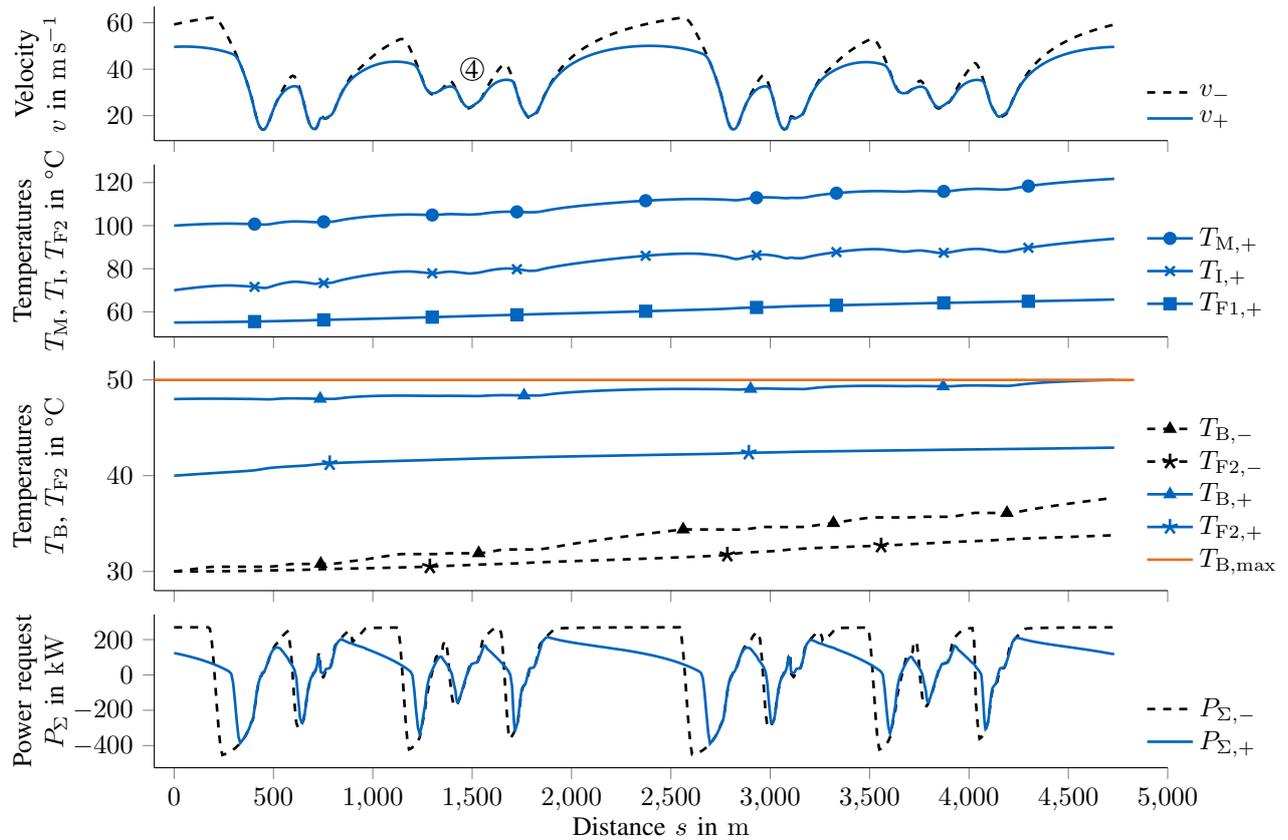}
	\caption{Optimal race strategies for two laps on the Monteblanco race circuit showing velocity $v$, power request $P_\Sigma$ and component temperatures $T_c$.}
	\label{fig:plot_vTP}
\end{figure*}

The results were calculated using an i7-7820HQ CPU and \SI{16}{\giga\byte} of memory. The \gls{NLP} was solved with the primal-dual interior-point method IPOPT interfaced by CasADi \cite{Andersson2018} using a direct collocation transcription. The execution time of the solver for the \gls{NLP} for two race laps was approximately \SI{2.5}{\minute}. The discretization step size varied along the race-track. In curves, a finer mesh was implemented to allow for a better description of the rapidly changing variables and their gradients leading to a step size of $\Delta s=\SI{3}{\meter}$. On the straight parts, a coarser mesh of $\Delta s=\SI{9}{\meter}$ was sufficient to reach high numerical precision in combination with small computation times. In total, approx. $44\cdot 10^3$ decision variables and $50\cdot 10^3$ constraints were present.\\
Two minimum race-time control strategies for a race, consisting of two laps, can be seen in Fig. \ref{fig:plot_vTP}. Two different cases were considered: In case ``cold'' ($-$), the initial temperatures of all the powertrain components $T_{c,0,-}$ equaled the environment temperature $T_\mathrm{env}$. In case ``hot'' ($+$), the initial component temperatures $T_{c,0,+}$ were set to the values given in Table \ref{tab:ini_vals}.
\begin{table}[h]
\caption{Initial temperature values $T_{c,0}$ of powertrain components.}
\label{tab:ini_vals}
\begin{center}
\renewcommand{\arraystretch}{1.3}
\begin{tabular}{rccccc}
& $T_{\mathrm{M},0}$ & $T_{\mathrm{I},0}$ & $T_{\mathrm{B},0}$ & $T_{\mathrm{F1},0}$ & $T_{\mathrm{F2},0}$\\
& \multicolumn{5}{c}{in \SI{}{\degreeCelsius}}\\
\hline
``cold'' ($-$) & 30 & 30 & 30 & 30 & 30\\
``hot'' ($+$) & 100 & 70 & 48 & 55 & 40\\
\end{tabular}
\end{center}
\end{table}\\
In case ``cold'', none of the components $c$ reached their maximum allowed temperature. Therefore, the maximum vehicle power of $P_\Sigma=Fv$ of \SI{270}{\kilo\watt} could be requested at all times when allowed by the driving dynamics as displayed in the last plot. The maximum physically achievable velocity of approx. $\SI{220}{\kilo\meter\per\hour}$ for this vehicle on the Monteblanco race-track resulted on the straights. The battery temperature $T_\mathrm{B}$ remained far below the limit of $T_\mathrm{B,max}=\SI{50}{\degreeCelsius}$.\\
Case ``hot'' shows the necessity and performance of the developed control-strategy. Here, the initial component temperatures $T_{c,0,+}$ were set to a valid combination of reasonable values (Table \ref{tab:ini_vals}) that can occur during a race. The optimization's initial battery temperature $T_{\mathrm{B},0,+}$ almost equaled $T_\mathrm{B,max}$. The race-strategy then was adapted to the given conditions to reach $T_\mathrm{B,max}$ exactly when crossing the finish line at $s=\SI{4.73}{\kilo\meter}$ to avoid a safety stop during the race. This was achieved by augmented phases of lift and coast in comparison with the race-strategy for case ``cold'': As the requested power $P_{\Sigma,+}$ shows, the vehicle braked later before curves. Additionally, the requested power $P_{\Sigma,+}$ slowly decreased on the straights. When the driving resistances could not be overcome by $P_{\Sigma,+}$, the vehicle's velocity $v_+$ decreased till the entry of the next curve. Therefore, the breaks could be applied late. Still, acceleration phases overlapped in both strategies, even if they ended in different maximum velocities $v$. The requested power maxima differed in their absolute values. The influence on the vehicle speed $v_+$ in case ``hot'' is evident: Physically, maximum velocity was never reached and mean acceleration as well as deceleration occurred less aggressively, meaning the velocity's gradients were decreased.\\
The coolant fluid temperature $T_\mathrm{F2,+}$ reached an equilibrium at the end of the optimization horizon since only as much heat was allowed to be released internally by the battery as the coolant fluid could dissipate. Coolant circuit $\mathrm{F1}$ could be neglected in this case. Machine and inverter temperatures $T_\mathrm{M,+}$ and $T_\mathrm{I,+}$ did not reach their limits.\\
Another important point is the difference in the race paths to be driven in both cases (Fig. \ref{fig:plot_path}). In case ``hot'' velocity $v_+$ in curve 4 (marked) was higher. This per se had a positive influence on the lap time. Nevertheless, the higher curve speed $v_+$ required the path to change slightly within and immediately after this turn. Since the combined tire usage was already at the limit here, the race-path radius in case ``hot'' increased. By this, its curvature decreased and the higher speed $v_+$ could feasibly be driven. Nevertheless, the decreased acceleration after turn 4 led to a diminished increase of the powertrain temperatures.\\
The total race times cumulated over the two laps were \SI{142.12}{\second} in case ``cold'' and \SI{149.59}{\second} in case ``hot''.
%%%%%%%%%%%%%%%%%%%%%%%%%%%%%%%%%%%%%%%%%%%%%%%%%%%%%%%%%%%%%%%%%%%%%%%%%%%%%%%%%%%%%%%%%%%%%%%%%%%%%%%%%%%%%%%
%\addtolength{\textheight}{-0.5cm}
\section{CONCLUSION \& OUTLOOK}
\label{sec:ConclusionOutlook}
In the next Roborace season, the methods presented in this publication will be applied to the racecar. On the one hand, the available energy can then be used as effectively as possible. On the other hand, the powertrain components can be exploited as much as possible without loosing race time. Additionally, powertrain losses and component temperatures can then be compared with measurement data we receive when driving the proposed global race lines that were calculated considering the thermodynamical influence.\\
Furthermore, an additional optimization will be implemented that allows for the mentioned online re-planning of the race-strategy in the presence of disturbances. With the help of the results in this publication this simplified online optimization can be realized.\\
Along with these improvements the loss-models will be replaced by physically more detailed descriptions of the powertrain components.
%\addtolength{\textheight}{-20cm}   % This command serves to balance the column lengths
                                  % on the last page of the document manually. It shortens
                                  % the textheight of the last page by a suitable amount.
                                  % This command does not take effect until the next page
                                  % so it should come on the page before the last. Make
                                  % sure that you do not shorten the textheight too much.

%%%%%%%%%%%%%%%%%%%%%%%%%%%%%%%%%%%%%%%%%%%%%%%%%%%%%%%%%%%%%%%%%%%%%%%%%%%%%%%%

%%%%%%%%%%%%%%%%%%%%%%%%%%%%%%%%%%%%%%%%%%%%%%%%%%%%%%%%%%%%%%%%%%%%%%%%%%%%%%%%

%%%%%%%%%%%%%%%%%%%%%%%%%%%%%%%%%%%%%%%%%%%%%%%%%%%%%%%%%%%%%%%%%%%%%%%%%%%%%%%%
\section*{CONTRIBUTIONS}
T. H. initiated the idea of the paper and contributed significantly to the concept, modeling, and results. F. P. modeled the powertrain thermodynamics in his Master’s thesis. J. B. contributed to the whole concept of the paper. M. L. provided a significant contribution to the concept of the research project. He revised the paper critically for important intellectual content. M. L. gave final approval for the publication of this version and is in agreement with all aspects of the work. As a guarantor, he accepts responsibility for the overall integrity of this paper.

\section*{ACKNOWLEDGMENT}
We would like to thank the Roborace team for giving us the opportunity to work with them and for the use of their vehicles for our research project. Special thanks to Ollie Walsh for generously sharing measurement data. We would also like to thank the Bavarian Research Foundation (Bayerische Forschungsstiftung) for funding us in connection with the ``rAIcing'' research project.

%\addtolength{\textheight}{20cm}

%%%%%%%%%%%%%%%%%%%%%%%%%%%%%%%%%%%%%%%%%%%%%%%%%%%%%%%%%%%%%%%%%%%%%%%%%%%%%%%%
%\dobeforekey{}{\newpage}
\bibliographystyle{IEEEtran}
\bibliography{IEEEabrv,referencesBIBTEX}

\end{document}